\documentclass[aps,preprint,floats,epsf,epsfig,nofootinbib,letter]{revtex4-1}

\usepackage{epsfig}

\begin{document}
\def\be{\begin{eqnarray}}
\def\en{\end{eqnarray}}
\def\non{\nonumber}
\def\la{\langle}
\def\ra{\rangle}
\def\nc{N_c^{\rm eff}}
\def\vp{\varepsilon}
\def\ep{\epsilon}
\def\drho{\bar\rho}
\def\deta{\bar\eta}
\def\A{{\cal A}}
\def\B{{\cal B}}
\def\c{{\cal C}}
\def\d{{\cal D}}
\def\e{{\cal E}}
\def\p{{\cal P}}
\def\t{{\cal T}}
\def\CP{{\it CP}~}
\def\up{\uparrow}
\def\dw{\downarrow}
\def\vma{{_{V-A}}}
\def\vpa{{_{V+A}}}
\def\smp{{_{S-P}}}
\def\spp{{_{S+P}}}
\def\J{{J/\psi}}
\def\ov{\overline}
\def\Lqcd{{\Lambda_{\rm QCD}}}
\def\pr{{Phys. Rev.}~}
\def\prl{{Phys. Rev. Lett.}~}
\def\pl{{Phys. Lett.}~}
\def\np{{Nucl. Phys.}~}
\def\zp{{Z. Phys.}~}
\def\lsim{ {\ \lower-1.2pt\vbox{\hbox{\rlap{$<$}\lower5pt\vbox{\hbox{$\sim$}
}}}\ } }
\def\gsim{ {\ \lower-1.2pt\vbox{\hbox{\rlap{$>$}\lower5pt\vbox{\hbox{$\sim$}
}}}\ } }

\begin{flushright}
{\small
CYCU-HEP-10-15 \\[0.2cm]
October, 2010}
\end{flushright}


\title{$B$ to Light Tensor Meson Form Factors Derived from Light-Cone Sum Rules}

\author{\sc Kwei-Chou Yang}\email{kcyang@cycu.edu.tw}

\affiliation{\vspace*{0.3cm} \normalsize\sl Department of Physics,
Chung Yuan Christian University, Chung-Li, Taiwan 320 \vspace*{2cm} }



\begin{abstract}
\vskip0.5cm
Using the recent results for the two-parton light-cone distribution amplitudes of the tensor meson, we calculate the form factors for the decays of $B_{u,d,s}$ into the light $J^{PC}=2^{++}$ tensor mesons via the vector/axial-vector/tensor current with the light-cone sum rules. We also obtain the $q^2$-dependence of the form factors.
\end{abstract}


\small

\maketitle


 {\bf 1.}~~
Semileptonic and radiative $B$ decays, which involve light tensor mesons and are related to $b \to s(d)$ transitions, contain rich phenomena relevant to the standard model and new physics.
It is interesting to note that due to the $G$-parity, both the chiral-even and chiral-odd two-parton light-cone distribution amplitudes (LCDAs) of the light $J^{PC}=2^{++}$ tensor meson are antisymmetric under the interchange of the momentum fractions of its $quark$ and $anti$-$quark$ in the SU(3) limit \cite{Cheng:2010hn}. Therefore, tensor mesons cannot be produced from the local $V\pm A$ currents of the standard model. The two-body $B$ decays involving tensor mesons are of great interest because these decays can further shed light on the underlying helicity structure \cite{cheng-yang-2010}.

In this paper we calculate the form factors for the $B_{u,d,s}$ decays into light $J^{PC}=2^{++}$ tensor mesons ($T$) via the vector/axial-vector/tensor current in the light-cone sum rule (LCSR) approach. These form factors are relevant to exclusive $B$ decays involving the tensor meson in the final state. Some model calculations for these form factors can be found in the literature \cite{ISGW,CCH,Hatanaka:2009gb,Hatanaka:2009sj,Wang:2010ni}.
In the quark model language, the $J^{PC}=2^{++}$ tensor meson is described by a constituent quark-antiquark pair with angular momentum $L=1$ and total spin $S=1$. The observed tensor mesons $f_2(1270)$, $f_2'(1525)$, $a_2(1320)$ and $K_2^*(1430)$ form an SU(3) $1\,^3P_2$ nonet. Although light-cone sum rules have been widely used to
calculate various form factors, $B\to$ tensor meson form factors have not been systematically studied \footnote{In \cite{Safir:2001cd}, the $T_1(0)$ form factor for the $B \to K_2^*$ transition was calculated by using the LCSR approach. However, the assumed leading-twist LCDA for the tensor meson does not match its G-parity property.}. The relevant inputs in this study are the LCDAs of the tensor mesons.  However, only chiral-even two-parton LCDAs for the $f_2^\prime (1525)$ were given about ten years ago \cite{Braun:2000cs}. Recently, we have systematically studied the chiral-even and chiral-odd two-parton LCDAs for $J^{PC}=2^{++}$ tensor mesons \cite{Cheng:2010hn}.


\vskip 0.7cm

{\bf 2.}~~ For a tensor meson with mass $m_T$ and four-momentum $(E, 0, 0, P_3)$ moving along the $z$-axis, its polarizations $\epsilon_{(\lambda)}^{\mu\nu}$ with helicity $\lambda$ can be represented in terms of the polarization vectors \cite{Berger:2000wt}
\begin{eqnarray}
\varepsilon(0)^{*\mu} = (P_3,0,0,E)/m_T,
\quad
\varepsilon(\pm1)^{*\mu} = (0,\mp1,+i,0)/\sqrt{2},
\end{eqnarray}and are given by
\begin{eqnarray}
\epsilon^{\mu\nu}_{(0)} &\equiv& \sqrt{\frac{1}{6}}
 [\varepsilon(+1)^\mu \varepsilon(-1)^\nu + \varepsilon(-1)^\mu \varepsilon(+1)^\nu]
 + \sqrt{\frac{2}{3}}  \varepsilon(0)^\mu \varepsilon(0)^\nu ,
 \\
\epsilon^{\mu\nu}_{(\pm1)} &\equiv& \sqrt{\frac{1}{2}}
[\varepsilon(\pm1)^\mu \varepsilon(0)^\nu + \varepsilon(0)^\mu \varepsilon(\pm1)^\nu], \quad
\epsilon^{\mu\nu}_{(\pm2)} \equiv \varepsilon(\pm1)^\mu \varepsilon(\pm1)^\nu .
\end{eqnarray}
The polarization tensors $\epsilon^{(\lambda)}_{\alpha\beta}$ are symmetric and
traceless. Moreover, they satisfy the divergence-free condition $\epsilon^{(\lambda)}_{\alpha\beta} P^\beta=0$ and the orthonormal condition $\epsilon^{(\lambda)}_{\mu\nu}\big(\epsilon^{(\lambda') \mu\nu}\big)^*=\delta_{\lambda\lambda'}$.

By considering the semileptonic $B$ decays involving tensor mesons, we can define the form factors: \cite{Hatanaka:2009gb,Hatanaka:2009sj,Wang:2010ni}
\begin{eqnarray} \label{eq:FF-def}
 \langle T(P, \lambda)|\bar{q}\gamma_\mu b|{ B} (p_B)\rangle
 &=& -i \frac{2}{m_B + m_{T}} \varepsilon_{\mu\nu\alpha\beta}
 e_{(\lambda)}^{*\nu}
 p_B^\alpha P^{\beta} V^{BT}(q^2),
 \\
 \langle T (P,\lambda)|\bar{q}\gamma_\mu \gamma_5 b|{ B}(p_B)\rangle
 &=& (m_B + m_{T}) e^{(\lambda)*}_{\mu} A_1^{BT}(q^2)
 - (e^{(\lambda)*} \cdot p_B)
(p_B + P)_\mu \frac{A_2^{BT}(q^2)}{m_B + m_{T}}
\nonumber \\
&& - 2 m_{T} \frac{\epsilon^{(\lambda)*}\cdot p_B}{q^2} q^\mu
\left[A_3^{BT}(q^2) - A_0^{BT}(q^2)\right],
\\
\langle T(P, \lambda)|\bar{q} \sigma^{\mu\nu}q_\nu b|{ B} (p_B)\rangle
&=&
 2  T_1^{BT} (q^2) \varepsilon^{\mu\nu\rho\sigma} p_{B\nu} P_{T\rho} e^*_\sigma,
\\
\langle T(P, \lambda)|\bar{q} \sigma^{\mu\nu}\gamma_5 q_\nu b|{ B} (p_B)\rangle
&=& -i T_2^{BT} (q^2) \left[
(m_B^2 - m_T^2) e^{*\mu} - (e^* \cdot p_B)(p_B^\mu + P_T^\mu)
\right]
\nonumber\\
 &&
- i T_3^{BT}(q^2) (e^* \cdot p_B)
\left[ q^\mu - \frac{q^2}{m_B^2 - m_T^2}(p_B^\mu + P_T^\mu)
\right],
\end{eqnarray}
where $q_\mu=(p_B-P)_\mu$ and $e^{*\mu}_{(\lambda)}\equiv \epsilon^{*\mu\nu}_{(\lambda)}q_{\nu}/m_B$.
We adopt the convention $\epsilon^{0123}=-1$, and
 \begin{eqnarray} \label{eq:a3}
  A_3^{BT}(q^2) &=&  \frac{m_B + m_{T}}{2 m_{T}} A_1^{BT}(q^2) - \frac{m_B
 - m_{T}}{2 m_{T}} A_2^{BT}(q^2).
 \end{eqnarray}
The {\em tensor} form factors can also be defined in the following way:
\begin{eqnarray}\label{eq:FF-o-1}
 & & \langle T (P, \lambda)  | \bar  q \, \sigma^{\mu \nu}\gamma_5  b
 |\bar B_q (P_B) \rangle  =
 -iA^{BT} (q^2) \,
  \left [ e_{(\lambda)}^{* \mu}\, (P + P_B)^\nu  -
  (P + P_B)^\mu \,  e_{(\lambda)}^{* \nu}\right ] \nonumber\\
 & & \qquad +i B^{BT} (q^2) \,
  \left [ e_{(\lambda)}^{* \mu}\, q^\nu  -
  q^\mu \,  e_{(\lambda)}^{* \nu}\right ]
 +2 i C^{BT} (q^2) \,
 \frac{e_{(\lambda)}^* q}{m_{B_q}^2-m_T^2} \,
 \left [ P^\mu q^\nu - q^\mu P^\nu \right ].
\end{eqnarray}
We then find that
\begin{eqnarray}
 T_1^{BT} (q^2) &=& A^{BT}(q^2),  \nonumber\\
 T_2^{BT}(q^2) &=& A^{BT}(q^2)- \frac{q^2}{m_{B_q}^2 -m_T^2} B^{BT}(q^2),  \nonumber\\
 T_3^{BT} (q^2) &=& B^{BT}(q^2) + C^{BT} (q^2).
\end{eqnarray}

\vskip 0.7cm

{\bf 3.}~~ To calculate the form factors, we consider the following two-point correlation functions, which are sandwiched between the vacuum and transversely polarized $T$ meson with $\lambda=\pm1$:
\begin{eqnarray}
&& i\int d^4x e^{iq x} \langle T(P,\lambda=1)|T [\bar
q_1(x)\gamma_\mu(1-\gamma_5) b(x)\, j_{B_{q_2}}^\dagger(0)]|0\rangle \nonumber\\
&& ~~ = -{\bf A_1}(p_B^2,q^2) e^{*(\perp)}_\mu
 + {\bf A_2}(p_B^2,q^2) (e^{*(\perp)} q) (2P+q)_\mu
 + {\bf A}(p_B^2,q^2) \frac{e^{*(\perp)} q}{q^2} q_\mu  \nonumber\\
&& ~~~~~\, -i {\bf V}(p_B^2,q^2)\varepsilon_{\mu\nu\rho\sigma} e^{*\nu}_{(\perp)} q^\rho P^\sigma \,,
\label{eq:correlator-1}
\end{eqnarray}
and
\begin{eqnarray}
\lefteqn{i\int d^4x e^{iq x} \langle T(P,\lambda=1)|T [\bar q_1(x)\sigma_{\mu\nu}\gamma_5
b(x)\, j_{B_{q_2}}^\dagger(0)]|0\rangle }
\hspace*{1.0cm} \nonumber\\
 & & = -i \mathcal{A}
 (p_B^2,q^2)
\{e^{*(\perp)}_\mu (2P+q)_\nu - e^{*(\perp)}_\nu (2P+q)_\mu\}
\nonumber\\
& & {}\ \  +i \mathcal{B}(p_B^2,q^2)\{e^{*(\perp)}_\mu q_\nu -
e^{*(\perp)}_\nu q_\mu\} + 2 i\mathcal{C} (p_B^2,q^2)
\,\frac{e^{*(\perp)} q}{m_{B_q}^2 -m_{T}^2}\, \{P_\mu q_\nu - q_\mu P_\nu\},
\label{eq:correlator-2}
\end{eqnarray}
where $$e^{*(\perp)}_\mu\equiv \frac{1}{\sqrt{2}}\varepsilon^*_\mu(\pm1) \frac{\varepsilon^*(0)\cdot q}{m_B},$$ $p_B^2=(P+q)^2$, $P$ is the momentum of the $T$ meson, and $j_{B_{q_2}}=i \bar
q_2 \gamma_5 b$ (with $q_{2(1)} \equiv u, d$ {\rm or} $s$) is the
interpolating current for the ${B_{q_2}}$ meson, so that
\begin{equation}\label{eq:B-decay-cont}
\langle 0|j_{B_{q_2}}(0)|\bar B_{q_2}(p_B)\rangle = \frac{f_{B_{q_2}} m_{B_{q_2}}^2}{m_b+m_{q_2}}\,.
\end{equation}

In the region of sufficiently large virtualities, i.e., $m_b^2 - p_B^2\gg
\Lambda_{\rm QCD}m_b$ where $q^2$ is small and positive, the operator product
expansion is applicable in (\ref{eq:correlator-1}) and (\ref{eq:correlator-2}), so that for an
energetic $T$ meson the correlation functions defined in (\ref{eq:correlator-1}) and (\ref{eq:correlator-2}) can
be generally represented in terms of the LCDAs of the $T$ meson (with $\Gamma \equiv$ the vector, axial-vector, or tensor current):
\begin{eqnarray}
\lefteqn{i\int d^4x e^{iq x} \langle T(P,\lambda=1)|T [\bar
q_1(x)\Gamma b(x)\, j_B^\dagger(0)]|0\rangle }
\hspace*{1.0cm} \nonumber\\
  & & = \int_0^1 du \frac{-i}{(q+ k)^2-m_b^2}
 {\rm Tr} \Big[\Gamma (\not\! q + \not\! k +m_b)\gamma_5
M_\perp^{T}(\lambda=1) \Big] \Bigg|_{k=uE n_-}
  +{\cal O}\bigg(\frac{m_{T}^2}{E^2}\bigg) \,,~~~\label{eq:green-fn-2}
\end{eqnarray}
where $E=|\vec{P}|$, $P^\mu =En_-^\mu + m_{T}^2 n_+^\mu/(4E) \simeq E n_-^\mu$
with two light-like vectors $n_-^\mu=(1,0,0,-1)$ and $n_+^\mu=(1,0,0,1)$
satisfying $n_- n_+=2$ and $n_-^2=n_+^2=0$. Here $M_\perp^{T}$ is the transverse projector which is discussed as follows. We assign the momentum of the $q_1$-quark in the tensor meson as
 \begin{eqnarray}
 k^\mu = u E n_-^\mu +k_\perp^\mu + \frac{k_\perp^2}{4 uE}n_+^\mu\,,
\end{eqnarray}
where $k_\perp$ is of order $\Lambda_{\rm QCD}$, $E$ is of order $m_b$, and $u$ is the momentum
fraction carried by the $q_1$-quark in the meson. In ({\ref{eq:green-fn-2}), to calculate contributions in the momentum space,
we use the following substitution (with the $k_\perp^2$ term omitted):
\begin{equation}\label{eq:x-to-k}
x^\mu \to -i \frac{\partial}{\partial k_{\mu}}\simeq -i \Bigg(
\frac{n_+^\mu}{2E}\frac{\partial}{\partial u} +
\frac{\partial}{\partial k_{\perp\, \mu}}\Bigg)\,,
\end{equation}
to the Fourier transform for
\begin{eqnarray}
 &&\langle T(P,\lambda)|\bar q^1_{\alpha}(x) \, q^2_\delta(0)|0\rangle
= -\frac{i}{4} \, \int_0^1 du \,  e^{i u  P x}
   \Bigg\{ f_T m_T^2 \Bigg[
    \not\!P \, \frac{\epsilon^{*(\lambda)}_{\mu\nu} x^\mu x^\nu}{(Px)^2} \,
    \Phi^T_\parallel(u) -\not\!x{\epsilon^{*(\lambda)}_{\mu\nu}x^\mu x^\nu \over 2(Px)^3} m_T^2\bar g_3(u)  \non \\
    && \qquad +  \left({\epsilon^{*(\lambda)}_{\mu\nu}x^\nu \over Px}-P_\mu{\epsilon^{*(\lambda)}_{\nu\beta}x^\nu z^\beta\over (Px)^2}\right)\gamma^\mu\, g_v(u)
     +  \frac{1}{2} \epsilon_{\mu\nu\rho\sigma}\gamma^\mu
    \epsilon^{*\nu\beta}_{(\lambda)} x_\beta P^\rho x^\sigma\gamma_5{1\over Px}\, g_a(u)  \Bigg] \non \\
  &&\qquad - \,{i\over 2}f^{\perp}_T m_T \Bigg[
  \frac{\sigma^{\mu\nu}
  \left(\epsilon^{(\lambda)*}_{\mu\beta} x^\beta P_\nu
 - \epsilon^{(\lambda)*}_{\nu\beta} x^\beta P_\mu\right)}{Px} \Phi^T_\perp(u)
   + \frac{\sigma^{\mu\nu}(P_\mu x_\nu - P_\nu x_\mu)
 m_T^2\epsilon^{(\lambda)*}_{\rho\beta}x^\rho x^\beta}{(Px)^3} \,\bar h_t(u) \nonumber  \\
 && \qquad +
  \sigma^{\mu\nu}\left(\epsilon^{(\lambda)*}_{\mu\beta} x^\beta x_\nu
 - \epsilon^{(\lambda)*}_{\nu\beta} x^\beta x_\mu\right) \frac{m_T^2}{2(Px)^2}\, \bar h_3(u)+ \epsilon^{*(\lambda)}_{\mu\nu}x^\mu x^\nu {m_T^2\over Px}\,h_s(u)\Bigg]+{\cal O}(x^2)\Bigg\}_{\delta\alpha}\,, \label{eq:DAs}
 \end{eqnarray}
where $x^2 \neq 0$ and we have
\begin{eqnarray}
 \bar g_3 (u) &=& g_3(u) +\Phi_\parallel^T -2 g_v(u),\nonumber\\
 \bar h_t(u) &=& h_t(u) - \frac{1}{2} \Phi_\perp^T(u) -\frac{1}{2} h_3(u) , \nonumber\\
 \bar h_3(u) &=& h_3(u) -\Phi_\perp^T(u).
\end{eqnarray}
$\Phi_\parallel^T, \Phi_\perp^T$ are of twist-2, $g_v, g_a, h_t, h_s$ are of twist-3, and $g_3, h_3$ are of twist-4. To leading-twist accuracy, we will use the following approximations for the relevant two-parton LCDAs in the present study \cite{Cheng:2010hn}:
\begin{eqnarray}\label{eq:DA}
\Phi_{\parallel,\perp}^T(u,\mu) &\simeq& 6u(1-u)
 a_1^{(\parallel,\perp),T}(\mu) C^{3/2}_1(2u-1) ,
\\
    g_v(u) &\simeq& \int\limits_{0}^u dv\, \frac{\Phi^T_\parallel(v)}{\bar v}+
                  \int\limits_{u}^1 dv\, \frac{\Phi^T_\parallel(v)}{v}\,,
\nonumber\\
    g_a(u) &\simeq& 2\bar{u}\int\limits_{0}^u dv\, \frac{\Phi^T_\parallel(v)}{\bar v}+
                  2u\int\limits_{u}^1 dv\, \frac{\Phi^T_\parallel(v)}{v}\,.
\end{eqnarray}
In general, the G-parity violating parameters $a_0^{(\parallel,\perp),T}$ vanish because the tensor meson cannot be produced from the local $V-A$ current. Consequently, from (\ref{eq:DAs}) we can obtain the light-cone projection for the $T$ meson in momentum space,
\begin{equation}
  M_{\delta\alpha}^T =  M_{\delta\alpha}{}_\parallel^T (\lambda=0)+
   M_{\delta\alpha}{}_\perp^T (\lambda=1) +
   M_{\delta\alpha}{}_\perp^T (\lambda=2) \,,
\label{eq:projector}
\end{equation}
where $M^T_{\delta\alpha}{}_\parallel$ and $M^T_{\delta\alpha}{}_\perp$ are
the longitudinal and transverse projectors, respectively. The transverse
projector $M_{\delta\alpha}{}_\perp^T (\lambda=1)$, which is relevant here,
is given by
 \begin{eqnarray} \label{eq:VproT}
 M^T_\perp(\lambda=\pm1) &=& -i\frac{f^{\perp}_T}{4} E\left[\epsilon^{*(\lambda)}_{\bot\mu\alpha}
  n_+^\alpha  \left(m_T\over 2E\right)\right]
 \Bigg\{
 \gamma^\mu\not\! n_- \,
   \Phi^T_\perp(u)\nonumber\\
& & +  \frac{f_T}{f_T^\perp} \frac{m_T}{E}
\Bigg[
 \gamma^\mu g_v(u) -  \, E\int_0^u dv\, \Big(2\Phi^T_\parallel(v) -g_v(v)\Big)\not\!n_-{\partial\over \partial k_{\bot\mu}} \non \\
  &-&  \,i \varepsilon_{\mu\nu\rho\sigma} \, \gamma^\nu n_-^\rho \gamma_5
  \left(n_+^\sigma \,{g'_a(u)\over 4}- E\,\frac{g_a(u)}{2} \,
  \frac{\partial}{\partial k_\perp{}_\sigma}\right)
 \Bigg]
 +{\cal O} \left({m_T^2\over E^2}\right) \Bigg\}\,,
\end{eqnarray}
where
\begin{eqnarray}
\epsilon^{*(\lambda)}_{\bot\mu\nu}
  n_+^\nu \left({m_T\over 2E}\right) &=& \sqrt{1\over 2}\,\varepsilon^*_\mu(\pm1)\delta_{\lambda,\pm1}\,.
\end{eqnarray}
From the expansion of $M_{\delta\alpha}{}_\perp^T (\lambda=1)$, one can find that the contributions arising from
$g_v, 2\Phi_\parallel^T -g_v$, and $g_a$ are suppressed by $m_{T}/E$ relative to the term with
$\Phi_\perp$, i.e., the expansion parameter in the light-cone sum rules should be $m_T/m_b$, rather than the twist. Note that in (\ref{eq:green-fn-2}) the derivative with respect to the transverse momentum acts on the hard scattering amplitude before the collinear approximation is taken.

At the quark-gluon level, after performing the integration of (\ref{eq:green-fn-2}), the results up to ${\cal O}(m_T/m_b)$ read
\begin{eqnarray}
 \bf{A_1}^{\rm QCD}&=&  \frac{m_B}{\varepsilon^*(0)\cdot q} \frac{m_b^2 f_{T}^\perp}{2} \int_0^1 \frac{du}{u}
\Bigg\{ \frac{1} {m_b^2 -up_B^2 -\bar u q^2}
  \Bigg[\frac{m_b^2 -q^2}{m_b^2} \Phi_\perp^T(u)
 +\Bigg( \frac{ m_{T}f_{T} }{m_b f_{T}^\perp } \Bigg)
  2 ug_v(u)
 \Bigg] \Bigg\} , \\
  \bf{A_2}^{\rm QCD}&=& \frac{m_B}{\varepsilon^*(0)\cdot q}\frac{m_b^2 f_{T}^\perp}{2} \int_0^1 du \Bigg\{
  \frac{1}
  {m_b^2 -up_B^2 -\bar u q^2} \frac{\Phi_\perp^T(u)}{m_b^2} +
  \frac{ m_{T}f_{T} }{ m_b f_{T}^\perp }
  \frac{2 \Phi_{tt}(u)}{(m_b^2-up_B^2 -\bar u q^2)^2}
   \Bigg\} , \\
  \bf{A}^{\rm QCD}&=& - q^2 \bf{A_2}^{\rm QCD} ,
\end{eqnarray}
\begin{eqnarray}
  \bf{V}^{\rm QCD}&=&  \frac{m_B}{\varepsilon^*(0)\cdot q} m_b^2 f_{T}^\perp \int_0^1 du \Bigg\{
  \frac{1}
  {m_b^2 -up_B^2 -\bar u q^2} \frac{\Phi_\perp^T(u)}{m_b^2} +
  \frac{ m_{T}f_{T} }{ m_b f_{T}^\perp }
  \frac{g_a(u)}{(m_b^2-up_B^2 -\bar u q^2)^2}
   \Bigg\} ,~~~~~
\end{eqnarray}
\begin{eqnarray}
  \mathcal{A}^{\rm QCD}&=&  \frac{m_B}{\varepsilon^*(0)\cdot q} \frac{m_b f_{T}^\perp}{2} \int_0^1 du \Bigg\{
  \frac{1}
  {m_b^2 -up_B^2 -\bar u q^2} \Bigg[ \Phi_\perp^T(u) +
  \frac{ m_{T}f_{T} }{m_b f_{T}^\perp }
  \Bigg(ug_v(u) + \Phi_{tt}(u) + \frac{g_a(u)}{2} \Bigg)\Bigg] \nonumber\\
  & & \ \ \ \ +
  \frac{ m_{T}f_{T} }{2 m_b f_{T}^\perp }
   \frac{(m_b^2+q^2)}{(m_b^2-up_B^2 -\bar u q^2)^2}g_a(u)
   \Bigg\} \,,
\end{eqnarray}
\begin{eqnarray}
 \mathcal{B}^{\rm QCD}&=& \frac{m_B}{\varepsilon^*(0)\cdot q} \frac{m_b f_{T}^\perp}{2} \int_0^1 du \Bigg\{
  \frac{1}
  {m_b^2 -up_B^2 -\bar u q^2} \Bigg[ \Phi_\perp^T(u) +
  \frac{ m_{T}f_{T} }{m_b f_{T}^\perp }
  \Bigg( (u-2)g_v(u) + \Phi_{tt}(u)
  \nonumber\\
 & &
   + \Big(\frac{1}{2}-\frac{1}{u}\Big) g_a(u) \Bigg)\Bigg] +
  \frac{ m_{T}f_{T} }{m_b f_{T}^\perp }
   \Bigg[ m_b^2 - (m_b^2 -q^2)\Bigg(\frac{1}{2}-\frac{1}{u}\Bigg) \Bigg] \frac{g_a(u)}{(m_b^2-up_B^2 -\bar u q^2)^2}
   \Bigg\} , ~~~~~
\end{eqnarray}
\begin{eqnarray}
  \mathcal{C}^{\rm QCD}&=& \frac{m_B}{\varepsilon^*(0)\cdot q} \frac{f_T m_T m_{B_{q_2}}^2}{2}
  \int_0^1 du \Bigg\{\frac{1}
  {(m_b^2 -up_B^2 -\bar u q^2)^2} \Bigg[2\Phi_{tt}(u) - g_a(u)\Bigg]  \Bigg\}\,,
\end{eqnarray}
where $\Phi_{tt}(u)\equiv \int_0^u dv\, (2\Phi_\parallel(v) - g_v(v))$
and $\bar u \equiv 1-u$.

As an example,  the form factor $A_1$ for the $B\to T$ transition can be
further approximated with the help of quark-hadron duality:
\begin{equation}
  A_1(q^2)\cdot \frac{m_{B_{q_2}}+m_T}{m^2_{B_{q_2}}-p_B^2} \cdot
        \frac{m_{B_{q_2}}^2 f_{B_{q_2}}}{m_b + m_{q_2}} = \frac{1}{\pi}\int_{m_b^2}^{s_0}
        \frac{{\rm Im}\mathbf{A_1}^{\rm QCD}(s,q^2)}{s-p_B^2}ds\,,
\end{equation}
where $s_0$ is the excited state threshold. After applying the Borel transform
$p_B^2 \to M^2$  \cite{SVZ} to the above equation,
we obtain
\begin{eqnarray}
 A_1(q^2) = \frac{(m_b + m_{q_2})}{(m_{B_{q_2}}+m_T) m_{B_{q_2}}^2 f_{B_{q_2}}}
 e^{m_{B_{q_2}}^2/M^2} \frac{1}{\pi}\int_{m_b^2}^{s_0}
       e^{s/M^2} {\rm Im}\mathbf{A_1}^{\rm QCD}(s,q^2) ds\,.
 \end{eqnarray}
We summarize the light-cone sum rule results as follows,
\begin{eqnarray}
 A^{{B_{q_2}}T}_1(q^2)
 &=& d \frac{(m_b + m_{q_2})m_b^2 f_{T}^\perp }{(m_{B_{q_2}}+m_T)m_{B_{q_2}}^2
 f_{B_{q_2}}} e^{(m_{B_{q_2}}^2 -m_b^2)/M^2}
 \nonumber\\
& & \times \int_0^1 \frac{du}{u} e^{\bar u(q^2 -m_b^2)/(uM^2)} \theta[c(u,s_0)]
 \Bigg[\Phi_\perp^T(u) \frac{m_b^2-q^2}{2u m_b^2}
 +\frac{m_{B_{q_2}}+m_T}{m_{B_{q_2}}}\frac{ m_{T}f_{T} }{m_b f_{T}^\perp }
 g_v(u) \Bigg] , ~~~
\end{eqnarray}
\begin{eqnarray}
 A^{{B_{q_2}}T}_2(q^2)
&=& d \frac{(m_b + m_{q_2}) (m_{B_{q_2}}+m_T) f_{T}^\perp }{2  m_{B_{q_2}}^2 f_{B_{q_2}}}
 e^{(m_{B_{q_2}}^2 -m_b^2)/M^2}
\int_0^1  \frac{du}{u} e^{\bar u(q^2 -m_b^2)/(uM^2)}
        \Bigg[ \Phi_\perp^T(u) \theta[c(u,s_0)]\nonumber\\
& & \ \ +  \frac{m_{B_{q_2}}}{m_{B_{q_2}}+m_T}\frac{2m_{T}m_b f_{T}}{uM^2 f_{T}^\perp } \Phi_{tt}(u)
 \Bigg(\theta[c(u,s_0)] + uM^2 \delta[c(u,s_0)] \Bigg) \Bigg]  ,
\end{eqnarray}
\begin{eqnarray}
 A_0^{{B_{q_2}}T}(q^2)
&=& A^{{B_{q_2}}T}_3(q^2) + d \frac{q^2 (m_b + m_{q_2}) f_{T}^\perp }
{4  m_{B_{q_2}}^2 m_T f_{B_{q_2}}}   e^{(m_{B_{q_2}}^2
-m_b^2)/M^2} \int_0^1 du \Bigg\{ \frac{1}{u} e^{\bar u(q^2 -m_b^2)/(uM^2)} \nonumber\\
& & \times        \Bigg[ \Phi_\perp^T(u) \theta[c(u,s_0)]
 +  \frac{2m_{T}m_b f_{T}}{uM^2 f_{T}^\perp } \Phi_{tt}(u)
 \Bigg(\theta[c(u,s_0)] + uM^2 \delta[c(u,s_0)] \Bigg) \Bigg] \Bigg\} ,~~~
\end{eqnarray}
\begin{eqnarray}
 V^{{B_{q_2}}T}(q^2)
 &=& d \frac{(m_b + m_{q_2}) (m_{B_{q_2}}+m_T) f_{T}^\perp }{2  m_{B_{q_2}}^2 f_{B_{q_2}}}
  e^{(m_{B_{q_2}}^2 -m_b^2)/M^2}
\int_0^1 \frac{du}{u} e^{\bar u(q^2 -m_b^2)/(uM^2)}
        \Bigg[ \Phi_\perp^T(u) \theta[c(u,s_0)]\nonumber\\
& & \ \ +
 \frac{m_{B_{q_2}}}{m_{B_{q_2}}+m_T}\frac{m_{T} m_b f_{T}}{uM^2 f_{T}^\perp } g_a(u)
 \Bigg(\theta[c(u,s_0)] + uM^2 \delta[c(u,s_0)] \Bigg) \Bigg] ,
\end{eqnarray}
\begin{eqnarray}
 T_1^{B_{q_2}T}(q^2) &=&
  d \frac{(m_b+m_{q_2})m_b f_{T}^\perp }{2 m_{B_{q_2}}^2 f_{B_{q_2}}}   e^{(m_{B_{q_2}}^2 -m_b^2)/M^2} \nonumber\\
 & & \times \int_0^1  \frac{du}{u} e^{\bar u(q^2 -m_b^2)/(uM^2)} \Bigg\{ \theta[c(u,s_0)] \Bigg[\Phi_\perp^T(u)
 +\frac{ m_{T}f_{T} }{m_b f_{T}^\perp } \Big(u g_v(u) + \Phi_{tt}(u)  + \frac{g_a(u)}{2}\Big) \Bigg] \nonumber\\
& & ~~ +  \frac{m_{T} f_{T}}{2 m_b f_{T}^\perp }
 (m_b^2+q^2)g_a(u)
 \Bigg( \frac{\theta[c(u,s_0)]}{uM^2} +  \delta[c(u,s_0)] \Bigg) \Bigg\},
\end{eqnarray}
\begin{eqnarray}
     T_2^{B_{q_2}T}(q^2)=T_1^{B_{q_2}T}(q^2) - \frac{q^2}{m_{B_{q_2}}^2 -m_{T}^2} \, B^{BT}(q^2),
 ~~~~T_3^{B_{q_2}T}(q^2) = B^{BT}(q^2) +C^{BT}(q^2) \,,
\end{eqnarray}
with
\begin{eqnarray}
 B^{BT}(q^2) &=& d\frac{(m_b+m_{q_2}) m_b f_{T}^\perp }{2 m_{B_{q_2}}^2 f_{B_{q_2}}}   e^{(m_{B_{q_2}}^2 -m_b^2)/M^2} \int_0^1   \frac{du}{u} e^{\bar u(q^2 -m_b^2)/(uM^2)} \Bigg\{\theta[c(u,s_0)] \Bigg[\Phi_\perp^T(u) +\frac{ m_{T}f_{T} }{m_b f_{T}^\perp }\nonumber\\
& &  \ \
  \times \Bigg( (u-2) g_v(u) + \Phi_{tt}(u)  + \Big(\frac{1}{2}-\frac{1}{u}\Big)  g_a(u)
   \Bigg) \Bigg] \nonumber\\
& &  +  \frac{m_{T} f_{T}}{m_b f_{T}^\perp }
  \Bigg[ m_b^2 - (m_b^2 -q^2)\Bigg(\frac{1}{2}-\frac{1}{u}\Bigg) \Bigg] g_a(u)
 \Bigg( \frac{\theta[c(u,s_0)]}{uM^2} +  \delta[c(u,s_0)] \Bigg)
  \Bigg\}, ~~~~~~ \label{eq:B-SR}
 \\
C^{BT}(q^2) &=&  d\frac{(m_b+m_{q_2}) m_{T} f_{T}}{2  f_{B_{q_2}}}   e^{(m_{B_{q_2}}^2 -m_b^2)/M^2} \nonumber\\
    && ~~~ \times \int_0^1 \frac{du}{u} e^{\bar u(q^2 -m_b^2)/(uM^2)} \Bigg(2\Phi_{tt}(u)
    -g_a(u) \Bigg)
 \Bigg( \frac{\theta[c(u,s_0)]}{uM^2} +  \delta[c(u,s_0)] \Bigg) \,, \label{eq:C-SR}
\end{eqnarray}
where $A^{{B_{q_2}}T}_3(q^2)$ is defined by (\ref{eq:a3}),
 $c(u,s_0)=us_0 -m_b^2 + (1-u) q^2$, and $$d =\frac{2m_T}{m_B} \frac{1}{1-q^2/m_B^2}.$$



\begin{table}[tbp!]
\caption{
Input parameters. $f^{(\perp)}_T(\mu)$ and $a^{(\parallel,\perp),T}_{1}(\mu)$ are
given at the scale $\mu=1\,\rm{GeV}$.
} \label{tab:input}
\begin{center}
\begin{tabular}{|c|c|c|c|c|c|c|c|}
\hline\hline
\multicolumn{8}{|c|}{Light tensor mesons \cite{Cheng:2010hn}}  \\
\hline
 \multicolumn{2}{|c|}{$T$}         & \multicolumn{2}{|c|}{$f_T~ [{\rm MeV}]$} &
 \multicolumn{2}{|c|}{$f_T^\bot~ [{\rm MeV}]$} & \multicolumn{2}{|c|}{$a_1^{\parallel,T},a_1^{\bot,T}$} \\
\hline
 \multicolumn{2}{|c|}{$f_2(1270)$} & \multicolumn{2}{|c|}{$102\pm6$} & \multicolumn{2}{|c|}{$117\pm25$} &
 \multicolumn{2}{|c|}{${5\over 3}$} \\
 \hline
 \multicolumn{2}{|c|}{$f'_2(1525)$}& \multicolumn{2}{|c|}{$126\pm4$} & \multicolumn{2}{|c|}{$65\pm12$} &
 \multicolumn{2}{|c|}{${5\over 3}$} \\
 \hline
 \multicolumn{2}{|c|}{$a_2(1320)$} & \multicolumn{2}{|c|}{$107\pm6$} & \multicolumn{2}{|c|}{$105\pm21$} &
 \multicolumn{2}{|c|}{${5\over 3}$} \\
 \hline
 \multicolumn{2}{|c|}{$K^*_2(1430)$} & \multicolumn{2}{|c|}{$118\pm5$} & \multicolumn{2}{|c|}{$77\pm14$} &
 \multicolumn{2}{|c|}{${5\over 3}$} \\
\hline
\hline
\multicolumn{8}{|c|}{Strange quark mass (GeV), pole $b$-quark mass (GeV), and couplings \cite{Yang:2008xw}} \\
\hline
\multicolumn{2}{|c|}{~~~~~$m_s(2\,\mbox{GeV})$~~~~~} &
\multicolumn{2}{|c|}{~~~~~~~$m_{b,pole}$~~~~~~~} &
\multicolumn{2}{|c|}{~~~~$\alpha_s(1~{\rm GeV})$~~~~} &
\multicolumn{2}{|c|}{$\alpha_s(2.2~{\rm GeV})$}
\\
\multicolumn{2}{|c|}{$0.09\pm 0.01$} &
\multicolumn{2}{|c|}{$4.85\pm0.05$} &
\multicolumn{2}{|c|}{$0.495$} &
\multicolumn{2}{|c|}{$0.287$}
\\
\hline
\hline
\multicolumn{8}{|c|}{{\it Effective} $B_{(s)}$ decay constants \cite{Khodjamirian:1998ji,Yang:2008xw}} \\
\hline
\multicolumn{4}{|c|}{$\bar{f}_{B}$  [MeV]} &
\multicolumn{4}{|c|}{$\bar{f}_{B_s}$ [MeV]}  \\
\multicolumn{4}{|c|}{$145\pm 10$} &
\multicolumn{4}{|c|}{$165\pm 10$}  \\
\hline
\hline
\end{tabular}
\end{center}
\end{table}


\vskip 0.7cm
{\bf 4.}~~ The wave functions of the isoscalar tensor states $f_2(1270)$ and $f'_2(1525)$ are
 \begin{eqnarray}
 f_2(1270) &=&
{1\over\sqrt{2}}(f_2^u+f_2^d)\cos\theta_{f_2} + f_2^s\sin\theta_{f_2} ~, \quad
 f'_2(1525) =
{1\over\sqrt{2}}(f_2^u+f_2^d)\sin\theta_{f_2} - f_2^s\cos\theta_{f_2} ~, \nonumber
 \end{eqnarray}
with $f_2^u\equiv u\bar u$ and likewise for $f_2^{d,s}$.  Because $\pi\pi$ is the dominant decay mode of $f_2(1270)$, and $f_2'(1525)$ decays are predominated by the $K\bar K$ mode (see Ref.~\cite{PDG}), the mixing angle should be small.  It was found that $\theta_{f_2}=7.8^\circ$ \cite{Li} and $(9\pm1)^\circ$ \cite{PDG}.  Therefore, we assume $f_2(1270)$ is primarily a $(u\bar u+d\bar d)/\sqrt{2}$ state, while $f'_2(1525)$ is predominated by $s\bar s$. We summarize the relevant parameters in Table \ref{tab:input}, where the $f_{f_2(1270)}$ and $f_{K_2^*(1430)}$ were also studied in \cite{Aliev:1981ju} and \cite{Aliev:2009nn}, respectively.
We neglect the possible corrections due to $m_u$ and $m_d$. The pole $b$ quark mass is used and the scale-dependent parameters are evaluated at the factorization scale $\mu_f=\sqrt{m_{B_q}^2-m_{b,pole}^2}$. In the numerical analysis for the form-factor sum rules, we choose the excited state threshold of the $B$ meson to be $s_0 = 35.5\pm 2.0$ GeV$^2$ and the Borel window $6.0$~GeV$^2< M^2 < 12.0$~GeV$^2$, which is consistent with my previous study \cite{Yang:2008xw}.
We use the {\it effective} $B$ decay constants
$\bar{f}_B=145\pm 10$~MeV and $\bar{f}_{B_s}\simeq 165\pm 11$~MeV \cite{Yang:2008xw}, which are obtained from the QCD sum rules of the $B$ decay constants without including the radiative corrections to the coefficient of the unity operator. The former is in accordance with \cite{Khodjamirian:1998ji}.  In \cite{Yang:2008xw}, we have
checked that, using the present values of $f_B$ and $m_b$ in the LCSR of $B\to \rho$ form factors of the same order, we can get results which are in good agreement
with those given in \cite{Ball:2004rg} where the radiative corrections are included.  It was found that the radiative corrections to form factors can be canceled if one
adopts the $f_B$ sum rule result with the same order of $\alpha_s$-corrections
in the calculation \cite{Ball:1998kk,Ball:2004rg}. Therefore, the radiative
corrections might be negligible, so we do not include them.

We also assume that this set of parameters can apply to form factors with various $q^2$. Our results from fitting the $q^2$-dependence of form factors within the range $0\leq q^2 \leq 6$~GeV$^2$ are exhibited
in Tables \ref{tab:FF-BT} and \ref{tab:FF-BsT}, where the $q^2$-dependence is parameterized in
the three-parameter form:
\begin{eqnarray} \label{eq:FFpara}
 F^{B_q T}(q^2)=\,{F^{B_q T}(0)\over 1-a(q^2/m_{B_q}^2)+b(q^2/m_{B_q}^2)^2}\,,
\end{eqnarray}
with $F^{B_q T}\equiv A^{B_q T}_{0,1,2}$, $V^{B_q T}$ or $T^{B_q T}_{1,2,3}$. For simplicity, we
do not show the theoretical errors for the parameters $a$ and $b$. The magnitude of $T_3^{B_q T}(0)$ is small in general due to the fact that $B^{B_q T}(0)$ and $C^{B_q T}(0)$, defined by (\ref{eq:B-SR}) and (\ref{eq:C-SR}), have similar magnitudes but opposite signs.

\begin{table}[t]
\caption{$B\to$ tensor meson form factors obtained in the LCSR calculation are fitted to
the 3-parameter form in (\ref{eq:FFpara}). The error for $F^{BT}(0)$ is due to the variations of the Borel mass, decay constants, strange quark mass, and pole $b$ quark mass, which are then added in quadrature.} \label{tab:FF-BT}
\begin{ruledtabular}
\begin{tabular}{ c| c c c }
~~~$F$~~~~~
    & $[F^{Ba_2}(0),a,b]$~~~~~
    & $[F^{BK_2^*}(0),a,b]$~~~~~
    & $[F^{Bf_2}(0),a,b]$~~~~~~
 \\
    \hline
$A_1$
    & [$0.14\pm0.02, 1.21, 0.52$]
    & [$0.14\pm0.02, 1.23, 0.49$]
    & [$0.14\pm0.02, 1.20, 0.54$]
    \\
$A_2$
    & [$0.09\pm0.02, 1.27, 2.04$]
    & [$0.05\pm0.02, 1.32, 14.9$]
    & [$0.10\pm0.02, 1.45, 1.58$]
    \\
$A_0$
    & [$0.21\pm0.04, 1.19,-0.26$]
    & [$0.25\pm0.04, 1.57, 0.10$]
    & [$0.20\pm0.04, 0.99,-0.34$]
    \\
$V$
    & [$0.18\pm0.02, 2.10, 1.50$]
    & [$0.16\pm0.02, 2.08, 1.50$]
    & [$0.18\pm0.02, 2.10, 1.51$]
    \\
$T_1$
    & [$0.15\pm0.02, 2.09, 1.50$]
    & [$0.14\pm0.02, 2.07, 1.50$]
    & [$0.15\pm0.02, 2.10, 1.50$]
    \\
$T_2$
    & [$0.15\pm0.02, 1.21, 0.39$]
    & [$0.14\pm0.02, 1.22, 0.35$]
    & [$0.14\pm0.02, 1.20, 0.41$]
    \\
$T_3$
    & [$0.04\pm0.02, 2.14, 23.6$]
    & [$0.01^{+0.02}_{-0.01},~~~~ 9.91, 276$]
    & [$0.06\pm0.02, 1.04, 6.36$]
    \\
\end{tabular}
\end{ruledtabular}
\end{table}
%

\begin{table}[t]
\caption{Same as Table \ref{tab:FF-BT} except for $B_s \to$ tensor meson transitions.} \label{tab:FF-BsT}
\begin{ruledtabular}
\begin{tabular}{ c| c c c }
~~~$F$~~~~~
    & $[F^{B_s K_2^*}(0),a,b]$~~~~~
    & $[F^{B_s f_2^\prime}(0),a,b]$~~~~~
 \\
    \hline
$A_1$
    & [$0.12\pm0.02, 1.23, 0.48$]
    & [$0.13\pm0.02, 1.25, 0.47$]
    \\
$A_2$
    & [$0.05\pm0.02, 1.32, 14.9$]
    & [$0.03\pm0.02, 4.71, 105$]
    \\
$A_0$
    & [$0.22\pm0.04, 1.57, 0.10$]
    & [$0.25\pm0.04, 1.72, 0.31$]
    \\
$V$
    & [$0.15\pm0.02, 2.08, 1.50$]
    & [$0.15\pm0.02, 2.06, 1.49$]
    \\
$T_1$
    & [$0.13\pm0.02, 2.07, 1.49$]
    & [$0.13\pm0.02, 2.06, 1.49$]
    \\
$T_2$
    & [$0.13\pm0.02, 1.22, 0.35$]
    & [$0.13\pm0.02, 1.23, 0.32$]
    \\
$T_3$
    & [$0.01^{+0.02}_{-0.01},~~~~ 9.91, 276$]
    & [$0.00^{+0.02}_{-0.00}$,~~ ---,~~ ---~~]
    \\
\end{tabular}
\end{ruledtabular}
\end{table}
%

\vskip 0.7cm
{\bf 5.} ~~In summary, using the recent analysis of tensor meson distribution amplitudes \cite{Cheng:2010hn}, we have calculated the form factors of $B$ decays into tensor mesons with the light-cone sum rule approach. We have fitted the $q^2$-dependence of the form factors in the range $0\leq q^2 \leq 6$~GeV$^2$. Owing to the $G$-parity, the two-parton light-cone distribution amplitudes of the tensor mesons are antisymmetric under the interchange of its quark and anti-quark momentum fractions in the SU(3) limit. The sum rule results for form factors are sensitive to the light-cone distribution amplitudes. The expansion parameter in the light-cone sum rules is $m_T/m_b$, rather than the twist. For the resulting sum rules, we have included the terms up to the order of $m_T/m_b$ in the
light-cone expansion. The results could be further improved with more precise parameters describing the distribution amplitudes and by including ${\cal O}(\alpha_s)$ corrections.

\section*{Acknowledgments}

I would like to thank T.~M.~Aliev for his correspondence.
This research was supported in part by the National Center for Theoretical Sciences and the
National Science Council of R.O.C. under Grant No. NSC99-2112-M-003-005-MY3.

%


\begin{thebibliography}{92}

\bibitem{Cheng:2010hn}
  H.~Y.~Cheng, Y.~Koike and K.~C.~Yang,
  Phys.\ Rev.\  D {\bf 82}, 054019 (2010).

\bibitem{cheng-yang-2010}
  H.~Y.~Cheng and K.~C.~Yang,
  arXiv:1010.3309 [hep-ph].


\bibitem{ISGW} N. Isgur, D. Scora, B. Grinstein, and M.B. Wise, Phys.\ Rev.\  D
{\bf 39}, 799 (1989).

\bibitem{CCH}
    H.~Y.~Cheng, C.~K.~Chua and C.~W.~Hwang,
    Phys.\ Rev.\  D {\bf 69}, 074025 (2004).

\bibitem{Hatanaka:2009gb}
  H.~Hatanaka and K.~C.~Yang,
  Phys.\ Rev.\  D {\bf 79}, 114008 (2009).

\bibitem{Hatanaka:2009sj}
  H.~Hatanaka and K.~C.~Yang,
  Eur.\ Phys.\ J.\  C {\bf 67}, 149 (2010).

\bibitem{Wang:2010ni}
  W.~Wang,
  arXiv:1008.5326 [hep-ph].



\bibitem{Safir:2001cd}
  A.~S.~Safir,
  Eur.\ Phys.\ J.\ direct C {\bf 3}, 15 (2001).

\bibitem{Braun:2000cs}
  V.~M.~Braun and N.~Kivel,
  Phys.\ Lett.\  B {\bf 501}, 48 (2001).

\bibitem{Berger:2000wt}
  E.~R.~Berger, A.~Donnachie, H.~G.~Dosch and O.~Nachtmann,
  Eur.\ Phys.\ J.\  C {\bf 14}, 673 (2000).


\bibitem{SVZ} M.A.\ Shifman, A.I.\ Vainshtein and V.I.\ Zakharov, Nucl.\
Phys.\ {\bf B147} (1979) 385, 448, 519.

\bibitem{PDG} K. Nakamura et al. (Particle Data Group), J. Phys. G 37, 075021 (2010).

\bibitem{Li} D.M. Li, H. Yu, and Q.X. Shen, J. Phys. G {\bf 27},
807 (2001)

\bibitem{Aliev:1981ju}
  T.~M.~Aliev and M.~A.~Shifman,
  Phys.\ Lett.\  B {\bf 112}, 401 (1982).

\bibitem{Aliev:2009nn}
  T.~M.~Aliev, K.~Azizi and V.~Bashiry,
  J.\ Phys.\ G {\bf 37}, 025001 (2010).

\bibitem{Yang:2008xw}
  K.~C.~Yang,
  Phys.\ Rev.\  D {\bf 78}, 034018 (2008).

\bibitem{Khodjamirian:1998ji}
  A.~Khodjamirian and R.~Ruckl,
  Adv.\ Ser.\ Direct.\ High Energy Phys.\  {\bf 15}, 345 (1998).

\bibitem{Ball:2004rg}
P.~Ball and R.~Zwicky,
  Phys.\ Rev.\  D {\bf 71}, 014029 (2005).

\bibitem{Ball:1998kk}
  P.~Ball and V.~M.~Braun,
  Phys.\ Rev.\  D {\bf 58}, 094016 (1998).




\end{thebibliography}
\end{document}